\documentclass[a4paper,11pt]{article}
\usepackage[normalem]{ulem}
\usepackage{mdframed,xcolor}
\usepackage{jheppub} 
\usepackage[T1]{fontenc} 

\usepackage{graphicx,epsfig}
\usepackage{times,bbm}
\usepackage{graphics,dcolumn,bm,float}
\usepackage{amssymb,amsmath,rotate,color,amsfonts}
\usepackage[title,titletoc,toc]{appendix}
\usepackage{mathtools}
\usepackage{booktabs}
\usepackage{tcolorbox}

\newtheorem{definition}{Definition}[section] 
\newtheorem{theorem}{Theorem}[section] 
\newtheorem{corollary}{Corollary}[section]

\usepackage{setspace}
\usepackage{wrapfig}
\usepackage{lipsum}
\usepackage{mwe}
\usepackage[mathlines]{lineno}
\usepackage[mathscr]{euscript}
\usepackage{breqn}
\usepackage{bbold}
\usepackage{hyperref}
\usepackage{pgfplots}
\usepackage{float}
\usepackage{tkz-euclide}
\usepackage{braket}
\usepackage{physics}
\usepackage[caption=false]{subfig}
\usepackage[export]{adjustbox}
\usepackage{tikz}
\usepackage{slashed}
\usepackage{bm}
\usepackage{multirow}
\usetikzlibrary{through,calc}
\usetikzlibrary{positioning}

\newcommand{\be}{\begin{equation}}
	\newcommand{\ee}{\end{equation}}
\newcommand{\bea}{\begin{eqnarray}}
	\newcommand{\eea}{\end{eqnarray}}
\newcommand{\nn}{\nonumber}

\title{\boldmath {Quantum Regular Black Holes and Complete Monotonicity}
}

\author[a]{A. Almasi,}
\author[b,1]{A. Moradpouri,\note{Corresponding author.}}
 \author[c]{and M. Shahbazi}

 \emailAdd{Ali.M.Almasi@gmail.com}
 \emailAdd{Ahmadreza.Moradpour@gmail.com}
 \emailAdd{Mojtaba.Shahbazi@modares.ac.ir}

\affiliation[a]{Department of Physics, Isfahan University of Technology (IUT), Isfahan 84156-83111, Iran.}
        \affiliation[b]{Research Center for High Energy Physics, Department of Physics, Sharif University of Technology, P.O.Box 11155-9161, Tehran, Iran.}
	\affiliation[c]{Department of Physics, School of Sciences,
	Tarbiat Modares University, Tehran, Iran.}

\abstract{
		We examine the conjecture for the complete monotonicity of certain curvature invariants for quantum black holes. In this note, we study a class of quantum regular black holes that are static, spherically symmetric, and characterized only by their mass. Additionally, this class of black holes reduces to the Schwarzschild solution in the classical limit $\hbar\to0$. We provide evidence supporting the non-perturbativity conjecture that perturbative corrections cannot falsify complete monotonicity. We demonstrate that these quantum black holes cannot be generated by perturbative  quantum corrections to the Einstein equations. We then investigate the thermodynamics of these black holes and derive a bound on their entropy, showing that the entropy is always greater than the horizon area divided by $4G$. We Also demonstrate that these black holes exhibit a bounded temperature, with a maximum temperature scaling as $T\sim\frac{1}{L_p}$ and a critical mass scale where the temperature vanishes.}

\begin{document} 
\maketitle
\flushbottom




\section{Introduction}
Quantum gravity is one of the primary goals of theoretical physics today. While several theories of quantum gravity have been proposed, such as string theory and loop quantum gravity, it is generally believed that no final theory has yet resolved all puzzles of gravity, such as the information paradox and dark energy, universally and satisfactorily. From the Bekenstein-Hawking entropy formula~\cite{Bekenstein:1972tm,Hawking:1975vcx}, it is reasonable to infer that the fundamental degrees of freedom of quantum gravity are not encoded in a pseudo-Riemannian metric; instead, the continuous description of spacetime is merely an approximation of a deeper theory, valid only at larger distances and in the weak-gravity limit. A key aspect of Einstein's theory of gravity is the singularity theorems, which assert that singularities are inevitable under certain reasonable conditions~\cite{PhysRevLett.14.57,Hawking:1967ju,PhysRevLett.90.151301}. For more examples, see~\cite{Senovilla_2015,Steinbauer_2022}, and for a review, see~\cite{Senovilla_1998}.  There is an ongoing debate over whether quantum or stringy effects can resolve the singularities of spacetime\footnote{String theory is consistent with certain kinds of singularities, such as orbifold singularities.} or not. It is generally believed that a hallmark of a good theory of quantum gravity is the absence of singularities; however, this is not universally accepted~\cite{Engelhardt:2016kqb}.  Fundamental objects of string theory at the perturbative level are one-dimensional strings rather than point particles, and this paradigm shift significantly alters the nature of spacetime. For example, T-duality predicts the existence of a minimal length scale in the toroidal compactification of string theory, constraining the curvature invariants of spacetime to be bounded. Furthermore, non-commutative geometries that arise in string theory, even at the closed string level, suggest that the curvature of spacetime cannot diverge arbitrarily and should be controlled by the string length scale~\cite{Nathan-Seiberg_1999,Freidel_2017}. Some evidence from cosmological models based on string theory~\cite{Brandenberger:1988aj,Gasperini_2003,gasperini2007elements} and loop quantum cosmology~\cite{Ashtekar_2011,Singh_2009} suggests that the big bang singularity can be avoided. In the context of black holes, there is also some compelling evidence from the approach of loop quantum gravity to the problem that the singularity of black holes may be resolved due to quantum corrections~\cite{ashtekar2023}. Hence, it is intriguing to study regular black holes even at the classical level, by incorporating higher-derivative curvature terms and different sources. The first model of a regular black hole was proposed by Bardeen~\cite{bardeen}, which was interpreted as a solution of the Einstein equations with a specific highly non-linear electromagnetic theory as its source ~\cite{Ay_n_Beato_2000}. Since then, extensive research has been conducted on regular black holes using different approaches. For examples, see~\cite{Dymnikova:1992ux,Ay_n_Beato_1998,Bronnikov_2001,Frolov_2016,Chamseddine_2017,Carballo_Rubio_2020,Borde_1997}, and for a review,~see~\cite{Lan_2023}. Two fundamental approaches to regularity are commonly employed: one is based on the finiteness of curvature invariants and the other on geodesic completeness, which requires that all null and timelike geodesics be complete, i.e., their affine parameter can be extended to infinity~\cite{Hawking_Ellis_1973,wald2010general}. These approaches are not generally equivalent; however, for static, spherically symmetric spacetimes where $g_{tt}g_{rr}=-1$, they are equivalent~\cite{Hu2023}.

Almost all attempts to construct regular black holes start with an action involving highly non-linear matter terms and possibly higher-derivative gravity terms. The foundational principles of such attempts remain unclear from a fundamental perspective. For example, the matter field proposed in~\cite{Ay_n_Beato_2000} to generate the Bardeen regular black hole is a highly non-linear electromagnetic model that is not connected to Maxwell's theory by changing the parameters of the theory. From a fundamental point of view, it is not clear how such a matter field can arise from first principles. Some attempts are based on computing perturbative quantum corrections to the Einstein-Hilbert action, which weaken the nature of singularities (for example, in some cases, the Kretschmann scalar diverges more slowly than in the classical Schwarzschild black hole~\cite{Duff,Calmet_2017,Bjerrum-Bohr:2002fji}); however, singularities remain, even after incorporating these perturbative corrections. The loop quantum gravity approach has attempted to calculate non-perturbative corrections to classical solutions, and in some cases has successfully resolved singularities~\cite{Modesto_2004,Gambini_2008,Perez_2017,bojowald2020,Alonso_Bardaji_2022,Alonso_Bardaji}.

However, since the nature of quantum gravity remains incompletely understood, we do not postulate any particular theory as the true quantum gravity theory to use as the starting point. Instead, we adopt a different approach and attempt to follow the fundamental principles expected to hold in any viable theory of quantum gravity. The crucial aspect of our approach is the no-singularity conjecture, which posits that a true theory of quantum gravity should be free of singularities. Furthermore, we assume that the closer one gets to the center of a black hole, the stronger the gravity becomes. The mathematical form of these assumptions is summarized as the complete monotonicity of certain curvature invariants, such as the Kretschmann scalar. Our last assumption is that regular black holes should reduce to the classical solutions of Einstein's theory of gravity in the limit $\hbar\to 0$. We will show that these assumptions significantly constrain the space of consistent solutions, and we will find a class of solutions that may serve as a foundation for further solutions.
 
The structure of this paper is as follows: In Section~\ref{general assumptions}, we introduce the basic assumptions and provide supporting evidence for them. Complete monotonicity is at the core of these considerations.Based on examples such as the quantum-corrected Schwarzschild black hole and different models of electrodynamics, we provide evidence for the non-perturbativity conjecture. This conjecture states that complete monotonicity cannot be falsified by perturbative corrections.  In Section~\ref{section3}, we delve into the mathematical definition of complete monotonicity and characterize a class of functions with the desired properties. Furthermore, since we do not postulate any action as a starting point, in Section~\ref{lastsection}, we define quantum black holes as configurations that satisfy the basic assumptions outlined in Section~\ref{general assumptions}. We then derive a large class of regular black holes that are consistent with the complete monotonicity of certain curvature invariants. We demonstrate that the black-hole singularities in static, spherically symmetric spacetimes cannot be resolved by perturbative corrections in $\hbar$ to the Einstein equations. We show that for a sufficiently large class of static, spherically symmetric spacetimes that are analytically characterizable, the entropy of quantum black holes satisfies the inequality $S>\frac{A_h}{4G}$, where $A_h$ is the radius of the horizon. It is conjectured that for all static, spherically symmetric spacetimes compatible with our assumptions, this bound cannot be violated.   

\section{General Assumptions}
{\label{general assumptions}}
In this section, we explore the general features that we believe any quantum analog of the Schwarzschild black hole should possess. As noted in Introduction, we expect that quantum gravity resolves any real singularity of spacetime, and the basic assumptions of this paper are as follows:
\begin{enumerate}
	\label{assump1}
	\item There is no real singularity at any point in spacetime.
\end{enumerate}
Another consistency check is that the solution(s) must reduce to the classical Schwarzschild black hole in the classical limit $\hbar\to0$. Thus, the second assumption is as follows:
\begin{enumerate}
	\label{assump2}
	\setcounter{enumi}{1}
	\item In the classical limit $\hbar\to0$, the solution(s) reduce to the classical Schwarzschild black hole.
\end{enumerate}
As we will discuss in the following sections, the meaning of the above assumption requires careful consideration. In fact, we will show that the components of the metric tensor are non-perturbative in $\hbar$, but the solution(s) exhibit a well-defined behavior in the limit $\hbar\to0$. The above premises are the most trivial starting point for finding regular black holes, but the third one is the most important and fundamental assumption of this paper.

To be clear, let us start with the following ansatz:
\bea
\label{schwartcoordinate}
ds^2=-(1-F(r))dt^2+\frac{1}{1-F(r)}dr^2+r^2d\Omega^2,
\eea
where for the classical Schwarzschild spacetime, $F(r)$ is $F_c(r)\equiv\frac{2M}{r}$. For $F(r)=F_c(r)$, the Kretschmann scalar and the expansion scalar for marginally bound timelike geodesics are given by:~\cite{Poisson:2009pwt,Carroll:2004st}
\bea
\label{krechmanneq}
K(r)=R_{\mu\nu\rho\sigma}R^{\mu\nu\rho\sigma}=\frac{48M^2}{r^6},~~~~~~\theta(r)=\pm\frac{3}{2}\sqrt{\frac{2M}{r^3}},
\eea
where $(+)$ corresponds to outgoing and $(-)$ corresponds to ingoing geodesics. As seen in Equation~\eqref{krechmanneq}, the above curvature invariants satisfy the following inequalities:
\bea
K(r+\lambda)<K(r),~~~~~~\theta(r+\lambda)<\theta(r),
\eea
for any positive real number $\lambda$. Indeed, the above curvature invariants satisfy even stronger constraints as follows:
\bea
(-1)^nK^{(n)}(r+\lambda)&<&(-1)^nK^{(n)}(r),\label{krechoutgoing}\\
(-1)^n\theta^{(n)}(r+\lambda)&<&(-1)^n\theta^{(n)}(r),\label{thetaoutgoing}
\eea
where $(n)$ denotes the number of derivatives with respect to $r$~(in~\eqref{thetaoutgoing}, outgoing geodesics is considered). The above behavior is the definition of a special class of functions known as \textit{completely monotonic} functions, which are defined as follows:
\begin{definition}
	Completely Monotonic Functions: An infinitely differentiable continuous function $f(r)$ is completely monotonic over $[0,\infty)$ if it satisfies:\footnote{We have slightly changed the definition, as in the literature $\geq$ is used instead of $>$. }
	\bea
	(-1)^n\frac{d^nf}{dr^n}>0.
	\eea
\end{definition}

All well-known examples of curvature invariants for the Schwarzschild spacetimes have the functionality $\sim 1/r^{\beta}$ for some positive $\beta$, which shows complete monotonicity. Intuitively, this is expected, since gravity strengthens as one moves toward the black hole. We will discuss the role of higher-derivative corrections to Einstein's gravity, which may have quantum or stringy origins, in more detail in the next sections.

Before going into more detail, let us focus on complete monotonicity. It is easy to find curvature invariants that are not completely monotonic.  Consider the following combination of the Kretschmann scalar $K$ and the Ricci tensor squared $R_{\mu\nu}R^{\mu\nu}$:
\bea
\label{noncmf}
L_1=R_{\mu\nu\rho\sigma}R^{\mu\nu\rho\sigma}-\beta R_{\mu\nu}R^{\mu\nu}.
\eea
It is possible to fine-tune $\beta$ in such a way that the above combination becomes zero at a coordinate distance $r=r^*$, violating the constraint $\mid L_1^{(n)}(r+\lambda)\mid<\mid L_1^{(n)}(r)\mid$. Another example is given easily by considering the classical Schwarzschild black hole and the curvature invariant $L_2=sin(R_{\mu\nu\rho\sigma}R^{\mu\nu\rho\sigma})$, which is highly oscillatory and deviates from all the inequalities near the origin. Thus, the inequalities~\eqref{krechoutgoing} and~\eqref{thetaoutgoing} do not hold for all curvature invariants.

Two important properties of the set of all completely monotonic functions are as follows:
\begin{itemize}
\item Closedness under addition with positive coefficients: If $f$ and $g$ are completely monotonic, then $af+bg$ is also a completely monotonic function for $a>0$ and $b>0$.
\item Closedness under multiplication: If $f$ and $g$ are completely monotonic functions, then $fg$ is also a completely monotonic function.
\end{itemize}

So, in general, the curvature invariant in Equation~\eqref{noncmf} is not expected to be a completely monotonic function for $\beta>0$, as it is a sum of two curvature invariants with negative coefficients. It forces us to consider curvature invariants that are candidates for complete monotonicity. 

The most natural candidates for complete monotonicity curvature invariants are:
\bea
K=R_{\mu\nu\rho\sigma}R^{\mu\nu\rho\sigma},~~~~~~~~\theta=\nabla_{\mu}u^{\mu}.
\eea  
where $u^\mu$ is the tangent vector field to a congruence of timelike geodesics. The Riemann tensor is the fundamental object that measures the curvature of spacetime, while the Kretschmann scalar quantifies the strength of curvature in an invariant manner, capturing the intuition that as one moves toward the center of a black hole, gravity becomes stronger. The expansion scalar has a clean geometrical meaning that quantifies the fractional rate of change of volume $\frac{1}{\delta V}\frac{d\delta V}{d\tau}$ for a congruence of timelike geodesics (and the fractional rate of change of cross-sectional area $\frac{1}{\delta A}\frac{d\delta A}{d\tau}$ for a congruence of null geodesics). The Raychaudhuri equation for a congruence of timelike geodesics in $3+1$ dimensions is given by:~\cite{Poisson:2009pwt}  
\bea
\frac{d\theta}{d\tau} = -\frac{1}{3} \theta^2 - \sigma^{\alpha\beta} \sigma_{\alpha\beta} + \omega^{\alpha\beta} \omega_{\alpha\beta} - R^{\alpha\beta} u_\alpha u_\beta,
\eea
which is the fundamental equation in exploring the singularity theorems and shows the importance of $\theta$ in analyzing singularities. Therefore, it is natural to consider $\theta$ as a second candidate for complete monotonicity. 

Although these curvature invariants may be good candidates for complete monotonicity, one cannot deduce that all other curvature invariants constructed from the Riemann tensor are expected to be completely monotonic. As an example, consider the trace-free part of the Riemann tensor, known as the Weyl tensor, and the trace-free part of the Ricci tensor, known as the traceless Ricci tensor, in $D$ dimensions $(D=d+1)$:
\bea
C_{\alpha\beta\gamma\delta}&=&R_{\alpha\beta\gamma\delta}-\frac{2}{D-2}(g_{\alpha[\gamma}R_{\delta]\beta}-g_{\beta[\gamma}R_{\delta]\alpha})+\frac{2}{(D-1)(D-2)}Rg_{\alpha[\gamma} g_{\delta]\beta},\nn\\
Z_{\mu\nu}&=&R_{\mu\nu}-\frac{1}{D}g_{\mu\nu}R,
\eea
which satisfy $g^{\alpha\gamma}C_{\alpha\beta\gamma\delta}=0$ and $g^{\mu\nu}Z_{\mu\nu}=0$. One can interpret the geometrical meaning of $C_{\alpha\beta\gamma\delta}$ and $Z_{\mu\nu}$ in two ways: On one hand, the vanishing of the Weyl tensor is associated with conformal flatness, while the vanishing of $Z_{\mu\nu}$
indicates that the manifold is an Einstein manifold, meaning~$R_{\mu\nu}=\lambda g_{\mu\nu}$ for some constant $\lambda$. On the other hand, $C_{\alpha\beta\gamma\delta}$ and $Z_{\mu\nu}$  are related to the evolution of the shear tensor~$\sigma_{\alpha\beta}$, which for a congruence of timelike geodesics in $3+1$ dimensions is:~\cite{Poisson:2009pwt}
\bea
\nabla_{\mu}\sigma_{\alpha\beta}u^{\mu}&=&-\frac{2}{3}\theta\sigma_{\alpha\beta}-\sigma_{\alpha\mu}\sigma^{\mu}_{\beta}-\omega_{\alpha\mu}\omega^{\mu}_{\beta}+\frac{1}{3}(\sigma^{\mu\nu}\sigma_{\mu\nu}-\omega^{\mu\nu}\omega_{\mu\nu})h_{\alpha\beta}-C_{\alpha\mu\beta\nu}u^{\mu}u^{\nu}\nn\\
&+&\frac{1}{2}Z^{TT}_{\alpha\beta},
\eea
where $h_{\alpha\beta}$ is the transverse metric $h_{\alpha\beta}=g_{\alpha\beta}+u_{\alpha}u_{\beta}$ and $Z^{TT}$ is the traceless part of the transverse Ricci tensor $R^T_{\alpha\beta}=h^{\alpha}_{\mu}h^{\beta}_{\nu}R_{\mu\nu}$. Hence, $C_{\alpha\beta\gamma\delta}$ and $Z_{\alpha\beta}$ are related to changes in the shape of the geodesic congruence. It is not clear whether their scalar invariants, $C_{\alpha \beta \gamma \delta}C^{\alpha \beta \gamma \delta}$ and $Z_{\mu \nu}Z^{\mu \nu}$ should be considered completely monotonic functions. In fact, it can be demonstrated that the complete monotonicity of the Kretschmann scalar contradicts the complete monotonicity of $Z_{\mu\nu}Z^{\mu\nu}$ for Ansatz~\eqref{schwartcoordinate}. Suppose that the Kretschmann scalar is a completely monotonic function. As shown in Subsection~\ref{quantumblackholes}, the core of the ansatz,~$r\sim 0$, is a de Sitter spacetime, which is an Einstein manifold. Therefore, $Z_{\mu\nu}=0$ near the center $(r\to 0)$. Moreover, since the ansatz should reduce to the Schwarzschild solution at large $r$, $Z_{\mu\nu}$ must be zero at this limit as well, which contradicts the complete monotonicity of $Z_{\mu\nu}Z^{\mu\nu}$. Furthermore, the de-Sitter spacetime is conformally flat, the Weyl tensor squared is not completely monotonic.

For our purposes, it suffices to assume that the Kretschmann scalar $K$ and the expansion scalar $\theta$ are completely monotonic functions. Therefore, the final assumption is as follows:
\begin{enumerate}
	\label{assump3}
	\setcounter{enumi}{2}
	\item  The Kretschmann scalar and the expansion scalar for marginally bound timelike geodesics are completely monotonic functions over $[0,\infty)$.
\end{enumerate}
Throughout this paper, we focus primarily on the complete monotonicity of $\theta$ for computational simplicity and subsequently verify the complete monotonicity of $K$.

\subsection{Evidence for Quantum Monotonicity}
Although the Kretschmann scalar and the expansion scalar for marginally bound timelike geodesics are completely monotonic functions for the classical Schwarzschild spacetime, it is instructive to explore how this complete monotonicity can be affected by higher-derivative terms and quantum corrections. If they remain completely monotonic, it provides more confidence that Assumption~\eqref{assump3} holds in the full quantum treatment of the Schwarzschild black hole. Unfortunately, there are some disagreements (or errors) in the results in the literature; for example, the results of~\cite{Nelson:2010ig} were criticized in~\cite{stelle}.  The most general quadratic action has the following form:~\cite{stelle}
\bea
\label{higherderivativeaction}
\mathscr{S}=\int d^4x\sqrt{-g}(\gamma R-\alpha C^{\mu\nu\rho\sigma}C_{\mu\nu\rho\sigma}+\beta R^2),
\eea
and the equations of motion (in the unit system $\gamma$=1) are given by:
\bea
\label{higherderivativeEOM}
R_{\mu\nu}-\frac{1}{2}g_{\mu\nu}R-4\alpha B_{\mu\nu}+2\beta R(R_{\mu\nu}-\frac{1}{4}g_{\mu\nu}R)+2\beta(g_{\mu\nu}\Box R-\nabla_{\mu}\nabla_{\nu}R)=0,
\eea
where $B_{\mu\nu}=(\nabla^{\rho}\nabla^{\sigma}+\frac{1}{2}R^{\rho\sigma})C_{\mu\rho\nu\sigma}$ is the Bach tensor, which is traceless. The coefficients $\alpha$ and $\beta$ depend on the parameters of microscopic theory, which may have quantum origins (necessitating an expansion in terms of $\hbar$) or other origins, such as $\alpha'$ corrections in string theory. Taking the trace of the above field equations leads to $\beta(\Box-\frac{1}{6\beta})R=0$. Using the techniques of~\cite{Nelson:2010ig}, one can show that for asymptotically flat, static, spherically symmetric spacetimes, one can set $R=0$. Using the Ricci scalar flatness condition $R=0$, the equations of motion~\eqref{higherderivativeEOM} simplify to:
\bea
\label{WeylEOM}
R_{\mu\nu}-\frac{1}{2}g_{\mu\nu}R=4\alpha B_{\mu\nu},
\eea
which are the equations of motion of the pure Weyl-Einstein gravity. The Ricci scalar flatness condition gives the following solution for Ansatz~\eqref{schwartcoordinate}:
\bea
F(r)=\frac{c_1}{r}+\frac{c_2}{r^2},
\eea
and the equations of motion~\eqref{WeylEOM} yield:~\cite{Silveravalle_2023}
\bea
&&4(rF'+F)+2r(rF''+2F')=0,\\
&&8\alpha r^2(1-F)^3F^{'2}+2\alpha r^3(1-F)^3F'F^{''}+4\alpha r^2(1-F)^4F^{''}\nn\\
&&+~4\alpha F'(1-F)^3-8\alpha(1-F)^4-4\alpha r F'(1-F)^4
+8\alpha(1-F)^5\nn\\
&&+~2r^2(1-F)^3(rF'+F)=0.
\eea
These equations imply that $c_2=0$. Therefore, the classical Schwarzschild black hole does not receive corrections from quadratic curvature terms, and complete monotonicity is preserved in the presence of higher-derivative terms. Intriguingly, there is some numerical evidence for another static, spherically symmetric spacetime beyond the standard Schwarzschild black hole~\cite{Huang_2023,stelle} that does not satisfy $g_{tt}g_{rr}=-1$, but since there is no analytic form for this solution, we do not consider it.

Quantum corrections to the Schwarzschild black hole have been computed in several papers ~\cite{Duff,Calmet_2017,Bjerrum-Bohr:2002fji}. The quantum-corrected Schwarzschild black hole in~\cite{Duff}, in the weak-gravity limit $\frac{2GM}{r}\ll 1$, has the following form: 
\bea
ds^2=-(1-\frac{L_s}{r}-\alpha\frac{L_sL_p^2}{r^3})dt^2+(1+\frac{L_s}{r}+\beta\frac{L_sL_p^2}{r^3})dr^2+r^2(1+\beta\frac{L_sL_p^2}{r^3})d\Omega^2,
\eea
where $L_s=2GM$ and $L_p$ is the Planck length. Its Kretschmann scalar up to $L_s^2L_p^2$ is given by:
\bea
\label{quantumKretchmann}
K(r)=\frac{12L_s^2}{r^6}+\frac{60(\alpha+\beta)}{r^8}L_s^2L_p^2.
\eea
To test complete monotonicity, we transform this quantity to the standard coordinates used in Ansatz~\eqref{schwartcoordinate}, yielding: 
\bea
r^2+\beta\frac{L_sL_p^2}{r}=\tilde r^2\Rightarrow r^2=\tilde r^2-\beta\frac{L_sL_p^2}{\tilde r}.
\eea
Therefore, the Kretschmann scalar~\eqref{quantumKretchmann} up to $(L_sL_p)^2$ is $K(r)=K(\tilde r)$. $\alpha$ and $\beta$ are numerical constants: $\alpha=13118\pi a$ and $ \beta=7752\pi a$ for some positive constant $a$~\cite{Duff}. Hence, $\alpha+\beta>0$, and the Kretschmann scalar remains completely monotonic under perturbative quantum corrections.

A more complete treatment of quantum corrections to the Schwarzschild black hole is discussed in~\cite{Bjerrum-Bohr:2002fji}, improving the result of~\cite{Duff} by considering the quantum effects of classical sources imposed in~\cite{Duff}. The corrected Schwarzschild black hole in the weak-field limit, ignoring terms such as $(\frac{L_s}{r})^2$ and higher orders and considering only first-order quantum correction terms proportional to $\frac{L_sL_p^2}{r^3}$, takes the following form:
\bea
\label{quantumSch}
ds^2=-(1-\frac{L_s}{r}+\frac{32L_sL_p^2}{15\pi r^3})dt^2+(1+\frac{L_s}{r}+\frac{45L_sL_p^2}{15\pi r^3})dr^2+r^2(1+\frac{L_s}{r}+\frac{7L_sL_p^2}{15\pi r^3})d\Omega^2.\nn\\
\eea
To convert the above solution to the standard coordinates used in Ansatz~\eqref{schwartcoordinate}, we use the following transformation (up to $(\frac{L_s}{r})^2$):
\bea 
r^2+L_sr+\frac{7L_sL_p^2}{15\pi r}=\tilde r^2\Rightarrow r^2=\tilde r^2-L_s\tilde r-\frac{7L_sL_p^2}{15\pi \tilde r}.
\eea
After straightforward calculations, we find:
\bea
\label{fullcorrectedSchw}
K(\tilde r)=K(r)=\frac{12L_s^2}{r^6}-\frac{160L_s^2L_p^2}{3\pi r^8}. 
\eea 
As seen, this is not a completely monotonic function; however, $K^{(n)}(r)$ changes sign around $r\sim L_p$, and the violation of the complete monotonicity of $K(r)$ occurs in a regime where perturbation theory is not valid. More precisely, perturbation theory only works in the limit $\frac{L_s}{r}\ll 1$, but for $r\sim L_p$, this limit is not satisfied. Therefore, the complete monotonicity of the Kretschmann scalar is not threatened by these considerations.
 
\subsection{Complete Monotonicity in Electrodynamics}
To gain a better understanding of complete monotonicity in quantum gravity, it is instructive to discuss this concept in electrodynamics. There is a major difference between the two: electrodynamics includes both positive and negative charges, whereas, in gravity, despite the existence of a negative Casimir energy, only particles with positive mass are known to exist. 

The Coulomb potential at tree level exhibits a completely monotonic behavior:
\bea
V(r)=-\frac{e^2}{4\pi r}.
\eea
Quantum corrections from vacuum polarization in $QED$ modify the Coulomb potential. At first order, the vacuum polarization diagram is: 
\begin{figure}[H]
	\centering
	\includegraphics[width=0.40\linewidth]{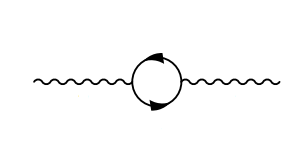}
	\caption{Vacuum polarization diagram at first order }
	\label{Feynman}
\end{figure}
\noindent and the one-loop corrected potential is given by:~\cite{schwartz2014quantum}
\bea
\label{Uhlingp}
V(r)=\frac{-e^2}{4\pi r}(1+\frac{e^2}{6\pi^2}\int_1^{\infty} dx e^{-2mrx}\frac{x^2+1}{2x^4}\sqrt{x^2-1}),
\eea
where its absolute value is clearly a completely monotonic potential.

The study of the finiteness of self-energy and electric field from a classical point of view has a long history, and many authors have discussed this issue ~\cite{Born:1934gh,Hale_2023,Kosyakov_2020,Tahamtan_2021}. One classical example of the finiteness of electric fields is the Born-Infeld electrodynamics~\cite{Born:1934gh}, where the Lagrangian is given by:\footnote{We have used the notations of~\cite{alam2022reviewborninfeldelectrodynamics}.}
\bea
\mathscr{L}=b^2(1-\sqrt{-det(\eta_{\mu\nu}+\frac{1}{b}F_{\mu\nu})}),
\eea
and the classical equations of motion without sources are:
\bea
\nabla.D&=&0,\nn\\
\frac{\partial D}{\partial t}&=&\nabla\times H,\nn\\
\nabla.B&=&0,\nn\\
-\frac{\partial B}{\partial t}&=&\nabla\times E,
\eea
where $D$ and $H$ are defined by $D=\epsilon E-\nu B$ and $H=\frac{1}{\mu}B-\nu E$. The parameters $\epsilon$, $\mu$, and $\nu$ are given by:
\bea 
\epsilon&=&\frac{1}{\sqrt{\Pi}},~~~~~~\mu=\sqrt{\Pi},~~~~~~\nu=\frac{\frac{1}{4}F_{\mu\nu}\tilde F^{\mu\nu}}{b^2\sqrt{\Pi}},
\eea
where $\Pi$ is defined by:
\bea
\Pi &=&1+\frac{1}{2b^2}(F_{\mu\nu} F^{\mu\nu})+\frac{1}{b^4}det(F_{\mu\nu}).
\eea
For a static charged point, applying $\nabla.D=0$ and $D=\frac{E}{\sqrt{1-\frac{E^2}{b^2}}}$ leads to the following solutions:
\bea
\label{full-nonperturnativeBI}
D(r)=\frac{q}{4\pi r^2},~~~~~~E(r)=\frac{b}{\sqrt{1+\frac{16\pi^2b^2}{q^2}r^4}},
\eea
where the electric field approaches a constant value at the position of the charged particle in a saturated manner, such that $\lim_{r\to 0}\frac{dE}{dr}= 0$, which is incompatible with complete monotonicity. 

There are plenty of other models of non-linear electrodynamics that regulate the singularity of the electric field of a point charge in different ways. We do not delve into the details and only focus on the results. In the regularized Maxwell theory~\cite{Tahamtan_2021,Hale_2023}, the electric field of a point charge is:
\bea
\label{regMAX}
E(r)=\frac{\alpha^2 Q}{(\alpha r+\sqrt{|Q|})^2},
\eea
where $\alpha$ and $Q$ are the parameters of the model. The electric field is completely monotonic. 

Another example of singularity-free non-linear electrodynamics is studied in ~\cite{Kosyakov_2020}. Although no analytic form of its $E(r)$ is available, the asymptotic form of the solution at the origin~$(r\to 0)$ exhibits complete monotonicity (at least up to its first derivative). One particular class of generalizations of Maxwell's theory is known as ModMax theory~\cite{Bandos_2020,Kosyakov_2020}, which focuses on conformal invariance and $SO(2)$ duality invariance in Maxwell's theory and finds the most general non-linear Hamiltonian that preserves these properties. Although point charges in models of this kind are singular, they exhibit complete monotonicity. 

For the last example, let us consider the quantum correction to Coulomb's law in scalar QED. The corrected Coulomb law has been computed to first order as follows~\cite{Helay_l_Neto_2000}:
\bea
\label{ScalarQED}
V(r)=-\frac{e^2}{r}(1-\frac{3}{64\pi^2}\frac{e^2L_c}{r}+\frac{5}{48\pi^4}\frac{e^2L^2_c}{r^2}),
\eea
where $L_c=\frac{\hbar }{mc}$ is the Compton wavelength. The above potential shows non-monotonic behavior if $V^{(n)}$ vanishes at some distance $r=r^*$, where $(n)$ is the number of derivatives with respect to $r$. The necessary condition to have a solution is the following condition:
\bea
e^2>\frac{2560(n+2)}{27(n+1)}.
\eea 
Again, as in the case of the corrected Schwarzschild solution~\eqref{fullcorrectedSchw} for the Kretschmann scalar, perturbation theory breaks down at such large coupling constants; therefore, the deviation from complete monotonicity cannot be trusted. Indeed, the long-distance behavior of the above solution is completely monotonic. 

As the above examples show, complete monotonicity is model-dependent. In the above examples, only Born-Infeld electrodynamics exhibits true non-monotonicity.

\subsection{The Non-perturbativity Conjecture}
In the previous subsections, we examined the role of quantum corrections in the Schwarzschild solution and their effects in different models of electrodynamics, focusing on how these corrections influence complete monotonicity. As shown, in both cases, there appears to be some degree of non-monotonicity. However, as we argued, this non-monotonicity cannot be trusted within the regime where perturbation theory is not valid. In fact, the long-distance behavior (i.e., complete monotonicity) of all corrected solutions is determined by the leading term, and the quantities of interest (such as electric field or curvature invariants) are, in fact, completely monotonic functions. This leads to the following conclusion, which we refer to as the non-perturbativity conjecture:
\begin{itemize}
	\label{non-complete}
	\item The Non-perturbativity Conjecture: Suppose that $f(r)$ is completely monotonic in the classical or weak-coupling limit. Then, perturbative corrections cannot invalidate complete monotonicity.
\end{itemize}
We provided some evidence supporting the validity of this conjecture for the corrected Schwarz-schild black hole~\eqref{fullcorrectedSchw} and for the Coulomb potential in scalar QED~\eqref{ScalarQED}. 

If we take this conjecture seriously, there are two possible outcomes for non-perturbative corrections to the desired quantities. The first possibility is that non-perturbative corrections could destroy complete monotonicity. An example of this type occurs in the potential between a quark and an antiquark, which, at short distances, is the Coulomb potential, but, at long distances (or in the strong-coupling limit), does not exhibit complete monotonicity. A commonly used model in the literature is the Cornell potential, given by $V(r) = kr - \frac{4\alpha_{QCD}}{3r}$, which is not a completely monotonic function. As a second example of this type, consider Born-Infeld electrodynamics and take $\frac{r_0}{r} \ll 1$, where $r_0^4 = \frac{q^2}{16\pi^2b^2}$, as the perturbation parameter. Then, the electric field of a point charge has the following first-order expansion:
\bea
E(r)=\frac{q}{4\pi}(\frac{1}{r^2}-\frac{q^2}{32\pi^2b^2r^6}).
\eea
Although the non-perturbativity conjecture holds for this expansion, and one cannot conclude from perturbation theory alone that it is not completely monotonic, the full non-perturbative Born-Infeld solution~\eqref{full-nonperturnativeBI} is not a completely monotonic function.

The second possibility is that complete monotonicity holds even at the non-perturbative level. As an example, consider the regularized Maxwell theory and take $\frac{\sqrt{|Q|}}{\alpha r}$ as the perturbation parameter. Then, the electric field~\eqref{regMAX} of a point charge takes the following form up to first order in perturbation theory:
\bea
E(r)=\frac{Q}{r^2}-\frac{2Q\sqrt{|Q|}}{\alpha r^3}.
\eea
If we ignore the non-perturbativity conjecture (which also holds in this case), the electric field is not a completely monotonic function. However, the full non-perturbative electric field~\eqref{regMAX} is a completely monotonic function.

The key assumption of this paper is to take the second possibility seriously in the context of quantum, regular, static, spherically symmetric black holes. Therefore, generalizing this property to the full non-perturbative level, at least for the mentioned curvature invariants, presents an intriguing problem and is worth exploring in the context of gravity.

\section{Completely Monotonic Functions}
\label{section3}
In this section, we summarize some important facts about completely monotonic functions that are very helpful in understanding the nature of this class of functions. Some examples of completely monotonic functions are provided below: 
\bea
\label{exampleofcm}
f_1(r)&=&e^{-\beta r} ~~~~~\text{for}~\beta>0,\nn\\
f_2(r)&=&\frac{1}{r^{\beta}} ~~~~~\text{for}~\beta>0,\nn\\
f_3(r)&=&\int_0^1e^{-zr}z^{\beta-1}dz ~~~~~\text{for}~\beta>0,
\eea
The unified nature of this class of functions is captured by Bernstein's theorem on completely monotonic functions \cite{schilling2012bernstein}:
\begin{theorem}
	\label{Berstein-th}
	Bernstein's Theorem: Let $f(r):[0,\infty)\to\mathbb{R}$ be a completely monotonic function. Then it is the Laplace transform of a unique measure $\mu$ on $[0,\infty)$ for all $r>0$:
	\bea
	f(r)=\mathcal{L}(\mu;r)=\int_0^{\infty}e^{-zr}\mu(dz).
	\eea
	Conversely, if $\mathcal{L}(\mu;r)<\infty$ for every $r>0$, then $r\to\mathcal{L}(\mu;r)$ is a completely monotonic function.   
\end{theorem}
  
 It is an extremely powerful theorem for characterizing completely monotonic functions and lies at the heart of all investigations concerning this class of functions. The measures corresponding to the first and second examples in~\eqref{exampleofcm} are $\mu(dz)=\delta(z-\beta)dz$ and $\mu(dz)=z^{\beta-1}dz$, respectively.
 
 Bernstein's theorem in its original form is not directly useful for our purposes. As shown in the previous sections, the Kretschmann scalar and the expansion scalar for marginally bound timelike geodesics in the classical Schwarzschild spacetime decay as $\sim\frac{1}{r^{\beta}}$ for some positive $\beta$; therefore, we need to specify conditions on the measure $\mu(dz)$ that guarantee that the corresponding monotonic function behaves asymptotically as $\sim\frac{1}{r^{\beta}}$. In the following corollary, we assume that $\mu(dz)=m(z)dz$, where $m(z)$ is a continuous, infinitely differentiable, positive function.

\subsection{Natural Measures}
{\label{Suitable function}}
In this subsection, we introduce a class of functions that are completely monotonic with the desired asymptotic behavior. The proposed functions, related to the lower incomplete gamma function, are as follows:
\bea
\label{suitablefunc}
M_{u,\alpha}(r)=\int_0^1e^{-zur}z^{\alpha-1}dz~~~~~~\text{for}~u>0~\text{and}~\alpha>0,
\eea 
which are completely monotonic. Clearly, for all $u>0$, in the limit $r\to\infty$, $M_{u,\alpha}(r)=\frac{\Gamma(\alpha)}{u^{\alpha}r^{\alpha}}$, where $\Gamma(\alpha)$ is the usual gamma function. The Bernstein representation of completely monotonic functions can be expressed in terms of $M_{u,\alpha}(r)$ functions as follows:
\bea
\label{alternativerep}
f(r)=(\alpha+r\frac{d}{dr})\int_0^{\infty}M_{u,\alpha}(r)\mu(du).\eea

To better understand Equation~\eqref{alternativerep}, we analyze the differential map $L_{\alpha}=\alpha+r\frac{d}{dr}$. The kernel of this map is the following class of functions:
\bea
k(r)=\frac{c^{*}}{r^{\alpha}},
\eea
where $c^{*}$ is a constant. However, it can be shown that this class of functions cannot be expressed as a combination of $M_{u,\alpha}(r)$. Assume that $k(r)$ has such a representation for a particular measure $\mu(du)$:
\bea
k(r)=\int_0^{\infty}M_{u,\alpha}(r)\mu(du)=\frac{1}{r^{\alpha}}\int_0^{\infty}\int_0^{ur}e^{-z}z^{\alpha-1}dz\frac{\mu(du)}{u^{\alpha}}\equiv\frac{J_\alpha(r)}{r^{\alpha}}.
\eea
Therefore, the derivative of $J_\alpha(r)$ with respect to $r$ should satisfy $\frac{dJ_\alpha(r)}{dr}=0$. However, this cannot hold. 

Let $\mathcal{S}(M_{\alpha})$ denote the space of all functions that admit a representation of the form $M_{\alpha}=\int_0^{\infty}M_{u,\alpha}(r)\mu(du)$ for some $\alpha$. The kernel of $L_{\alpha}$ for this domain is empty. Let $\mathcal{S}(CM)$ denote the space of all completely monotonic functions. Fixing $\alpha$, the map provided by:
\bea
\label{map}
L_{\alpha}: \mathcal{S}(M_{\alpha})\to \mathcal{S}(CM)
\eea 
is a bijection between the two spaces. If the bounded measure $\mu(du)$ satisfies the boundedness condition for the measure $\frac{\mu(du)}{u^{\alpha}}$, then it can be shown that the leading term in the limit $r\to\infty$ is proportional to $\frac{1}{r^{\alpha}}$. The map~\eqref{map} is much more interesting and leads to the following corollary:
 \begin{corollary}
	The necessary and sufficient condition for a completely monotonic function $f(r)$ over $[0,\infty)$ to be expressed in terms of $M_{u,\alpha}(r)$ is that its corresponding measure $\nu(du)=m_{\nu}(u)du$ induces a new measure $\mu(du)=m_{\mu}(u) du$, where $m_{\mu}(u)=(\alpha-1)m_{\nu}(u)-um'_{\nu}(u)$ for $\alpha>0$ and satisfies the asymptotic limits $\lim_{u\to0}um_{\nu}(u)=0$ and $\lim_{u\to\infty}um_{\nu}(u)e^{-ur}=0$. More precisely:
	\bea
	\label{corollary3.1}
	f(r)=\int_0^{\infty}e^{-ur}m_{\nu}(u)du=\int_0^{\infty}M_{u,\alpha}(r)m_{\mu}(u)du.
	\eea
\end{corollary}

 Proof: Suppose that $f(r)$ has such a representation:
 \bea
 f(r)=\int_0^{\infty}e^{-ur}m_{\nu}(u)du=\int_0^{\infty}M_{u,\alpha}(r)m_{\mu}(u)du.
 \eea
Applying $L_{\alpha}$ to both sides of the above equation yields:
\bea
\int_0^{\infty}e^{-ur}\Big((\alpha-1)m_{\nu}(u)-um'_{\nu}(u)\Big)du+um_{\nu}e^{-ur}|_0^{\infty}=\int_0^{\infty}e^{-ur}m_{\mu}(u)du.
\eea
The second term arises from integration by parts and vanishes due to the asymptotic limits. The boundary limits are expected, because the Laplace transform of $m_{\nu}$ must be finite. By Lerch's theorem~\cite{davies2012integral}, the Laplace transform is one-to-one; so we have:
\bea
\label{mu-measure}
(\alpha-1)m_{\nu}(u)-um'_{\nu}(u)=m_{\mu}(u).
\eea 
 For the converse, suppose that  $m_{\mu}(u)=(\alpha-1)m_{\nu}(u)-um'_{\nu}(u)$ is a measure. Then, $m_{\nu}(u)$ can be expressed in terms of $m_{\mu}(u)$ as follows:
 \bea
 m_{\nu}(u)=-u^{\alpha-1}\int_c^{u}m_{\mu}(y)y^{-\alpha}dy,
 \eea
where $c$ is an integration constant. Using integration by parts, $f(r)$ can be expressed as:
\bea
f(r)=-\int_0^{u}e^{-rz}z^{\alpha-1}dz\int_c^{u}m_{\mu}(y)y^{-\alpha}dy|_0^{\infty}+\int_0^{\infty} M_{u,\alpha}(r)m_{\mu}(u)du.
\eea
By choosing $c=\infty$, the boundary terms vanish, yielding the desired form for $f(r)$.

However, not all measures $m_{\nu}(u)$ lead to a measure such as $m_{\mu}(u)$. For example, consider $m_{\nu}(u)=1+sin(u)$. Here, no $\alpha$ exists such that $m_{\mu}(u)$ is a measure, since $\frac{um'_{\nu}(u)}{m_{\nu}(u)}$ does not attain a supremum. Hence, for a measure where $\frac{um'_{\nu}(u)}{m_{\nu}(u)}$ has a supremum, we obtain the following representation:
	\bea
	\label{SchwSolution}
		f(r)=\int_0^{\infty}M_{u,\alpha}(r)\mu(du).
	\eea
Assuming $0<\int_0^{\infty}\frac{\mu(du)}{u^{\alpha}}<\infty$, the asymptotic behavior of $f(r)$ is $\sim\frac{1}{r^{\alpha}}$ as $r\to\infty$. Equivalently, suppose that $g(r)$ is a completely monotonic function corresponding to the measure $m_{\nu}(u)$; if $h(r)=\alpha g(r)+rg'(r)$ is also a completely monotonic function for some $\alpha$, then $g(r)$ can be expressed as~\eqref{SchwSolution}. The representation~\eqref{SchwSolution} is not the most general form for a completely monotonic function with asymptotic behavior $\sim\frac{1}{r^{\beta}}$ as $r\to\infty$.

According to the Tauberian theorems~\cite{feller1968introduction}, if a completely monotonic function behaves as $\sim\frac{1}{r^{\beta}}$ as $r\to\infty$, its corresponding measure behaves as $\sim u^{\beta-1}$ as $u\to 0$. Therefore, to have the asymptotic behavior $\frac{1}{r^{\beta}}$ as $r\to\infty$, it suffices that the measure scales as $m_{\nu}(u)\sim u^{\beta-1}$ as $u\to 0$. Thus, let $m_{\nu}(u)=u^{\beta-1}n_{\nu}(u)$, where $n_{\nu}(u)\sim 1~$ as $~u\to 0$; then, Equation~\eqref{mu-measure} takes the following form:
\bea
\label{mu-measure2}
u^{\beta-1}\Big((\alpha-\beta)n_{\nu}(u)-un'_{\nu}(u)\Big)=m_{\mu}(u),
\eea 
and the function $n_{\nu}(u)$ does not need to satisfy $\frac{un'_{\nu}(u)}{n_{\nu}(u)}$ having a supremum. So, Equation~\eqref{SchwSolution} is not the most general form of a function with asymptotic behavior $\sim\frac{1}{r^{\beta}}$ as $r\to\infty$.

However, we expect that the subspace of measures where $\frac{um'_{\nu}(u)}{m_{\nu}(u)}$ attains a supremum suffices for our purposes. Assume that the set of points where $m_{\nu}=0$ has Lebesgue measure zero. Excluding this set from $u \in (0, \infty)$ leaves the left-hand side of Equation~\eqref{corollary3.1} unchanged. Thus, for simplicity, we assume that $m_{\nu}(u)$ is a strictly positive function. It can be shown that over an interval $[a,b] \subset (0,\infty)$, continuity and differentiability suffice to ensure that the supremum of $\frac{um'_{\nu}(u)}{m_{\nu}(u)}$ is finite. Since $m_{\nu}(u)$ is positive and continuous, $\frac{1}{m_{\nu}(u)}$ is also positive and continuous. Furthermore, the infinite differentiability of $m_{\nu}(u)$ ensures that $m'_{\nu}(u)$ is continuous and bounded. Consequently, $\frac{um'_{\nu}(u)}{m_{\nu}(u)}$ is continuous and bounded over $[a,b]$. By the extreme value theorem, there exist $c_1, c_2 \in [a,b]$ such that:
\bea
\frac{c_1m'_{\nu}(c_1)}{m_{\nu}(c_1)}\leq\frac{um'_{\nu}(u)}{m_{\nu}(u)}\leq\frac{c_2m'_{\nu}(c_2)}{m_{\nu}(c_2)}~~~~~~~\text{for all}~u\in[a,b].
\eea 

However, as mentioned earlier, this does not guarantee the existence of a supremum over $(0,\infty)$. This motivates defining a class of measures as natural measures as follows:
\begin{definition}
Natural Measures: $\nu(du) = m_{\nu}du$ is termed a natural measure if the existence of a supremum for $\frac{um'_{\nu}(u)}{m_{\nu}(u)}$ over any interval $[a,b]$ ensures its existence over $(0, \infty)$.
\end{definition}

In the next section, we focus on completely monotonic functions whose corresponding measures are natural.

\section{Quantum Black Holes}
\label{lastsection}
In this section, we focus on black hole configurations that are consistent with our assumptions. We denote these as quantum black holes, where the only scale beyond the mass of the black holes is the Planck length. Notably, we do not postulate any action as a starting point or quantize the degrees of freedom associated with such an action. For clarity, we define quantum black holes as follows:

\begin{definition}
	Quantum black holes:  A configuration $g_{\mu\nu}$ is called a quantum black hole, if it satisfies the three basic assumptions outlined in Section~\ref{general assumptions}, and if the Planck length is the only mass scale other than the mass of the black hole.
\end{definition} 

 The concept of complete monotonicity plays a crucial role in deriving such configurations. In the following subsections, we study the consequences of our assumptions, including the non-perturbative nature of singularity avoidance and the thermodynamic properties of these configurations. We show that the entropy of quantum black holes in such configurations satisfies $S>\frac{A_h}{4G}$, where $A_h$ is the area of the horizon. We explore the mass-temperature diagram for this class of regular black holes, showing that they exhibit a maximum temperature scaling as $T\sim\frac{1}{L_p}$,  in contrast to the classical Schwarzschild black hole, which is unstable under Hawking radiation and has a temperature that diverges.  Furthermore, we conjecture that this may be a fundamental feature of most spherically symmetric solutions in quantum gravity.

\subsection{Quantum Schwarzschild Black Hole}
\label{quantumblackholes}
After reviewing some basic facts about completely monotonic functions, we are ready to derive more quantitative results regarding Ansatz~\eqref{schwartcoordinate}. To proceed, we reformulate our assumptions from Section~\ref{general assumptions} in terms of complete monotonicity. The no-singularity assumption yields the following premise:
\begin{enumerate}
	\label{premise1}
	\item The measures are bounded:\footnote{Boundedness is defined over $r\in[0,\infty)$. For some measures, such as $\mu(du)=e^{-u}du$, a singularity occurs at a negative $r$, specifically $r=-1$ in this example. If such a singularity occurs, its physical meaning should be discussed.}
    \bea
    0<\int_0^{\infty}\mu(du)<\infty
    \eea
\end{enumerate}
The assumption regarding the classical limit of Ansatz~\ref{assump2} yields the following premises:
\begin{enumerate}
	\label{premise2}
	\setcounter{enumi}{1}
	\item The Kretschmann scalar and the expansion scalar for marginally bound timelike geodesics (denoted collectively as $N(r)$) in the Schwarzschild spacetime can be expressed as follows:\footnote{The asymptotic behavior of the Kretschmann scalar and the expansion scalar for marginally bound timelike geodesics in the Schwarzschild spacetime follows $\sim\frac{1}{r^{\beta}}$ for $\beta>1$.} 
	
	\bea
	\label{naturalmeasures}
	N(r)=\int_0^{\infty}M_{u,\alpha}(r)\mu(du).
	\eea 
\end{enumerate}
Since Assumption~\ref{assump3} is encompassed by the above two premises, we do not rephrase it separately.

The Kretschmann scalar for Ansatz~\eqref{schwartcoordinate} is given by:
\bea
K(r)=\frac{4F(r)^2}{r^4}+\frac{4F'(r)^2}{r^2}+F''(r)^2.
\eea
The regularity of $K(r)$ at the origin implies that the function $F(r)$ approaches zero at least quadratically as $r\to0$. Furthermore, to be compatible with complete monotonicity, $K(r)$ must approach zero exactly quadratically as $r\to0$. Therefore, $F(r)$ in the limit $r\to0$ is:
\bea
F(r)\sim r^2 +O(r^3),
\eea
which shows that the core of ansatz must be a de-Sitter spacetime. The asymptotic behavior of $F(r)$ as $r\to\infty$ must agree with the classical Schwarzschild spacetime; thus, in this region, it must equal $F_c(r)\equiv\frac{2M}{r}$. Collectively, the form of $F(r)$ in these limits is as follows:
\begin{equation}
	\label{assymptoticlimit}
	 F(r) \sim \{ \begin{array}{ll}
	ar^2+O(r^3) & \mbox{$r \to 0$}\\
	\frac{2M}{r} & \mbox{$r\to\infty$}.\end{array} 
 \end{equation}
The regularity of other curvature invariants is also consistent with the above asymptotic limits. For example, the expansion scalar $\theta$ and the Ricci scalar $R$ for Ansatz~\eqref{schwartcoordinate} are given by:~\cite{Poisson:2009pwt}
\bea
\label{KandTETA}
\theta(r)&=&\frac{(r^2\sqrt{F})'}{r^2},\\
R(r)&=&\frac{2F}{r^2}+\frac{4F'}{r}+F^{''},
\eea
whose regularity is compatible with the asymptotic behavior of $F(r)$ in Equation~\eqref{assymptoticlimit}.

\subsection{An Immediate Consequence: Non-perturbative Resolution of Singularities in $\hslash$}
The continuity of the function $F(r)$ over $[0,\infty)$ in addition to its asymptotic behavior given in Equation~\eqref{assymptoticlimit} implies that $F(r)$, roughly takes a schematic form similar to Figure~\ref{f(r)Fig}~(possibly with more than one critical point over $(0,\infty)$):
\begin{figure}[H]
	\centering
	\includegraphics[width=0.70\linewidth]{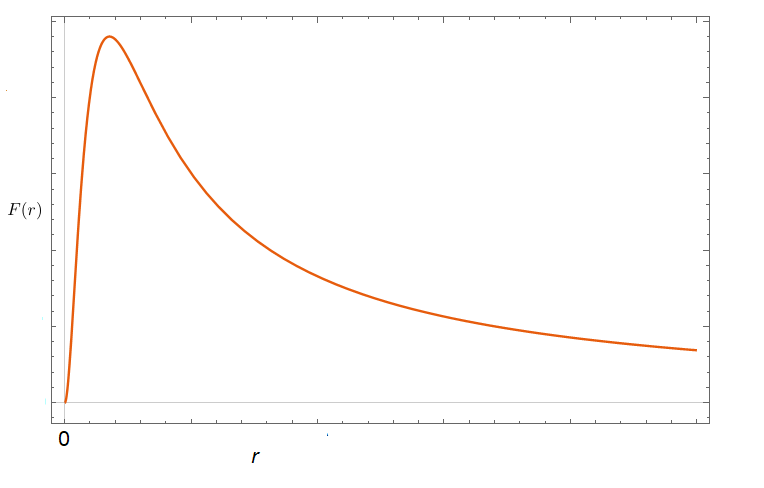}
	\caption{Schematic form of the function $F(r)$ }
	\label{f(r)Fig}
\end{figure}

Without loss of generality, in the following discussion, we assume that $F(r)$ has a single peak at $r=r^{*}$, where the maximum value of the curve is $F(r^{*})$. The natural scale in the quantum Schwarzschild black hole is the Planck length $L_p$; thus, we set $r^{*}\sim L_p$. The maximum value $F(r^{*})$ is then controlled by the Planck length and diverges in the classical limit $L_p\to 0$, as the solution must asymptotically satisfy $F(r)\sim \frac{1}{r}$ in this regime. Since the exact analytical form of this divergence is not crucial here, we assume that $F(r^{*})$ diverges as $F(r^{*})\sim\frac{1}{L_p}$. However, since the solution is regular, $F(r)$ near the center should behave as $F(r)\sim r^2$. Therefore,

in the region $r\in(0,r^{*})$, the difference $\frac{1}{2GM}|F(r)-F_c(r)|$ grows as $L_p\to0$,
where $F_c(r)\equiv\frac{2GM}{r}$ corresponds to the classical solution.~\footnote{Since the Planck length is explicitly present in these considerations, we recover G-dependence in all formulas.} Let us focus on the $F(r)$ function in the Bardeen model, expressed as follows~\cite{bardeen}:\footnote{In the original Bardeen black hole, $L_p$ on the right-hand side of Equation~\eqref{bardeenBH} is interpreted in~\cite{Ay_n_Beato_2000} as the magnetic monopole charge arising from highly nonlinear electrodynamics. For our work, it is instead interpreted as the Planck length.} 
\bea
\label{bardeenBH}
\frac{1}{2GM}F(r)=\frac{r^2}{(r^2+L_p^2)^{\frac{3}{2}}}.
\eea

The above function has the following expansions in two different regimes: 

\begin{equation}
	\label{assymptoticlimitOFbardeen1}
	\frac{1}{2GM}F(r)=
	\begin{cases}
		\dfrac{1}{r}\left(1 - \dfrac{3}{2}\left(\dfrac{L_p}{r}\right)^2 + \dfrac{15}{8}\left(\dfrac{L_p}{r}\right)^4 - \dots \right) & \text{$r > L_p$} \\[1.5ex] 
		\dfrac{r^2}{L_p^3}\left(1 - \dfrac{3}{2}\left(\dfrac{r}{L_p}\right)^2 + \dfrac{15}{8}\left(\dfrac{r}{L_p}\right)^4 - \dots \right) & \text{$r < L_p$}.
	\end{cases}
\end{equation}
The $L_p$ dependence in the second regime is non-perturbative in $L_p$. If we consider $\frac{r}{L_p}$ to be constant and take the limit $L_p \to 0$, $\frac{1}{2GM}F(r)$ diverges as $\frac{1}{L_p}$. This indicates that, in the regime $r < L_p$, non-perturbative effects must be accounted for to regularize the singularity of the classical black hole.

From a different perspective, the space of regular black holes is characterized by functions $F(r)$ that satisfy the boundary conditions $F(r=0) = 0$ (quadratic behavior near zero) and $F(r=\infty) = 0$. This space of functions cannot be continuously deformed (or perturbatively in $L_p$) into the space of functions with different boundary conditions, such as $F(r=0) = \infty$ and $F(r=\infty) = 0$, which are characteristic of the classical solution.

To explore this more quantitatively, we take one of the functions $M_{u,\alpha}$ as a test function. The scalar expansion for marginally bound timelike geodesics in the classical Schwarzschild spacetime behaves as $\sim\frac{1}{r^{\frac{3}{2}}}$~\cite{Poisson:2009pwt}; thus, for a test function, we choose $\alpha=\frac{3}{2}$ in Equation~\eqref{suitablefunc}:
\bea
\label{qtheta}
\theta(r)=\frac{\tilde{b}}{(ur)^{\frac{3}{2}}}\int_0^{ur}e^{-z}z^{\frac{1}{2}}dz,
\eea
where $\tilde{b}$ is a constant to be determined. This is related to a Dirac delta function as a measure in Equation~\eqref{naturalmeasures}. In the limit $r\to{\infty}$, $\theta (r)$ must match the classical expansion scalar in Equation~\eqref{krechmanneq}, leading to:
\bea
\frac{\tilde{b}}{u^{\frac{3}{2}}}=\frac{3\sqrt{2GM}}{\sqrt{\pi}}.
\eea
Since $u$ has a mass dimension, we can set $\frac{1}{u}=L_p\times\tilde{C}(\tilde{a})$, where $\tilde{a}\equiv\frac{L_p}{L_s}$ is a dimensionless parameter with $L_s\equiv2GM$, and $\tilde{C}(\tilde{a})$ is an arbitrary function of $\tilde{a}$. In the classical limit $L_p\to 0$, the upper bound in the integral in Equation~\eqref{qtheta} must go to infinity to match the classical behavior. The limit $\lim_{\tilde{a}\to 0}\tilde{a}\tilde{C}(\tilde{a})=0$ implies that $\tilde{C}(\tilde{a})$ must be at least of the form $\frac{1}{\tilde{a}^{1-\epsilon}}$ for some positive number $\epsilon$. For simplicity, we assume $\tilde{C}(\tilde{a})=1$. Thus, using Equation~\eqref{KandTETA} and Equation~\eqref{qtheta}, the form of the function $F(r)$ is given by:
\bea
\label{testfunction}
F(r)=\frac{18GM}{\pi}\frac{1}{r^4}\Big(\int_0^rdy\sqrt{y}\int_0^{\frac{y}{L_p}}e^{-z}z^{\frac{1}{2}}dz\Big)^2.
\eea  
We define $F(0)=0$ as $\lim_{r\to 0}{F(r)=0}$.

\begin{figure}[H]
	\centering
	\includegraphics[width=0.46\linewidth]{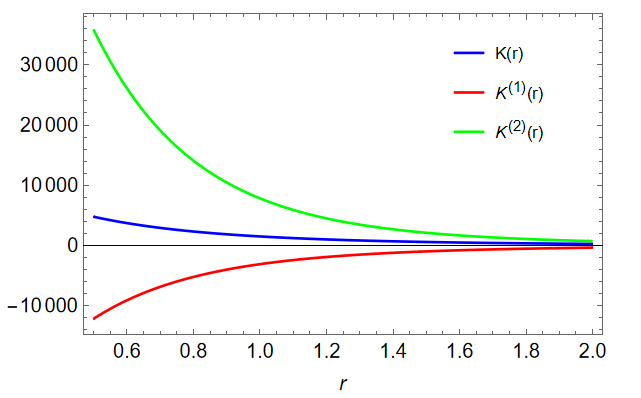}
	\includegraphics[width=0.46\linewidth]{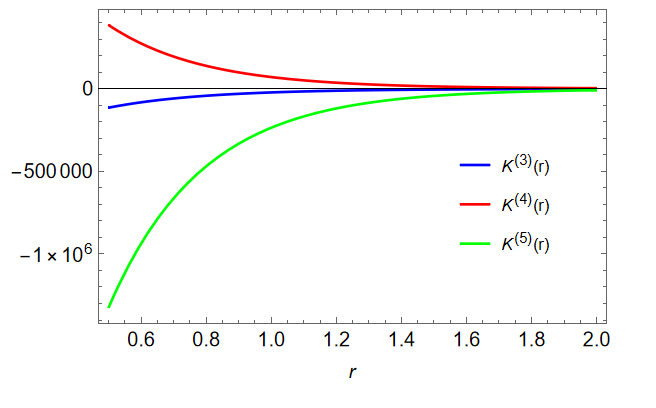}
	\caption{The Kretschmann scalar up to the fifth derivative. We have set $M=100$ and $L_p=1$.}
	\label{K}
\end{figure}

To better understand this, we plot $F(r)$ for different values of the Planck length~$L_p$ in Figure~\ref{diffrentPlancklength}:
\begin{figure}[H]
	\centering
	\includegraphics[width=0.73\linewidth]{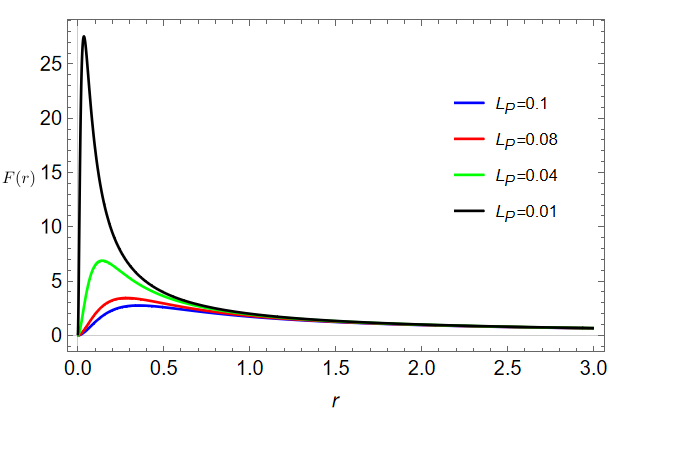}
	\caption{The function $F(r)$ for different values of the Planck length. We have set $M=1$. }
	\label{diffrentPlancklength}
\end{figure}
As is evident, the solution increasingly approximates the classical Schwarzschild spacetime over almost the entire half-line $(0,\infty)$, as $L_p$ increasingly approaches zero. Furthermore, it does not have a Taylor expansion in terms of $L_{p}$, as explicitly shown for the test function ~\eqref{testfunction}. We must also verify the complete monotonicity of the Kretschmann scalar. While direct verification is challenging, we have confirmed it up to the twelfth derivative. The complete monotonicity of the Kretschmann scalar is illustrated in Figure~\ref{K} up to the fifth derivative.\footnote{An effective approach to demonstrating complete monotonicity is through Bernstein's theorem, which requires proving that the Laplace inverse of the Kretschmann scalar is a positive function over $(0,\infty)$. However, due to numerical complexities, we have directly verified the complete monotonicity up to a certain order in derivatives.}

In the following subsections, we explore the thermodynamic properties of these regular black holes.

\subsection{Thermodynamic Properties of the Regular Schwarzschild Black Hole}
\label{entropyconsideration}
Exploring the thermodynamic properties of black holes is crucial, and regular black holes, rooted in quantum gravity, are no exception. We investigate how complete monotonicity and regularity influence their thermodynamic properties. We assume that Ansatz \eqref{schwartcoordinate} describes a black hole: 
\bea
\label{Schilsmetric}
ds^2=-(1-F(r))dt^2+\frac{1}{(1-F(r))}dr^2+r^2d\Omega^2.
\eea 

For now, we focus on the function $F(r)$ given by Equation~\eqref{testfunction}, and later generalize the results to a more generic case in Equation~\eqref{SchwSolution}. This ansatz describes a black hole if there exists a solution to $F(r_h)=1$. For $F(r)$ given by Equation~\eqref{testfunction}, the equation is:
\bea
\label{horizon}
1=\frac{18GM}{\pi}\frac{1}{r_h^4}\Big(\int_0^{r_h}dy\sqrt{y}\int_0^{\frac{y}{L_p}}e^{-z}z^{\frac{1}{2}}dz\Big)^2,
\eea
where $r_h$ is the radius of the horizon. As is evident from Figure~\ref{diffrentPlancklength}, the above equation typically has two solutions,~$r_{+}$ and $r_{-}$. For a critical mass of the order of the Planck mass~($M\sim m_p$), there is only a single solution. Otherwise, no black hole solution exists. In this subsection, we assume that the ansatz describes a black hole with $r_{+}\neq r_{-}$. From Equation~\eqref{horizon}, we obtain:
\bea
dM=\frac{\pi}{18G}d\Big(r_h^4(\int_0^{r_h}dy\sqrt{y}\int_0^{\frac{y}{L_p}}e^{-z}z^{\frac{1}{2}}dz)^{-2}\Big)\equiv\frac{\pi}{18G}dB(r_h).
\eea
The temperature of the black hole is given by $T=\frac{\kappa}{2\pi}$, where $\kappa$ is the surface gravity of the horizon. Thus, the temperature is:
\bea
T=\frac{1}{4\pi}|\frac{dF}{dr}|_{r=r_{+}}=\frac{1}{4\pi}\frac{dlog(B(r_h))}{dr_h}.
\eea
Using the first law of thermodynamics, the entropy of the black hole is as follows:
\bea
dS=\frac{dM}{T}=\frac{2\pi^2}{9G}B(r_h)dr_h.
\eea
Comparing this with the Bekenstein-Hawking formula $S=\frac{A}{4G}$ and defining a surface at a constant $r$, with area:\footnote{In higher-derivative theories, entropy is typically defined by the Wald formula~\cite{Wald_1993}, which differs from the simple Bekenstein-Hawking formula, area over $4G$. However, since we do not assume a specific higher-derivative Lagrangian, we directly apply the first law of thermodynamics to calculate entropy.} 
\bea
\label{qmSurface}
A=\frac{8\pi^2}{9}\int_0^{r_h}B(x)dx=\frac{8\pi^2}{9}\int_0^{r_h}\frac{x^4}{(\int_0^{x}dy\sqrt{y}\int_0^{\frac{y}{L_p}}e^{-z}z^{\frac{1}{2}}dz)^{2}}dx,
\eea
we recover the classical result $A=4\pi{r_h}^2$ in the limit $L_{p}\to 0$. Interestingly, the surface defined with the area $A$ in Equation~\eqref{qmSurface} is always greater than $4\pi{r_h}^2$, suggesting that this bound may serve as a criterion for true quantum Schwarzschild black holes.

Now, we examine this bound for a more general ansatz as given in Equation~\eqref{SchwSolution}. We take $\theta(r)$ as: 
\bea
\theta(r)=\tilde{c}\int_0^{\infty}\mu(du)M_{u,\frac{3}{2}}(r)=\frac{\tilde{c}}{r^{\frac{3}{2}}}\int_0^{\infty}\frac{\mu(du)}{u^{\frac{3}{2}}}\int_0^{ur}e^{-z}z^{\frac{1}{2}}dz,
\eea
where $\tilde{c}$ is a constant number and $u$ has a mass dimension. Rescaling $u$ as $u=\frac{w}{L_p}$ and defining $\tilde\mu(dw)$ as $\tilde\mu(dw)=\mu(du)$, we normalize the measure so that $\int_0^{\infty}\frac{\tilde\mu(dw)}{w^{\frac{3}{2}}}=1$.
We take the limit $r\to\infty$ such that $ur\to\infty$, which holds for $u\neq0 $, and we also assume this property for the case $u\to0$. Using similar reasoning, to ensure consistency with the asymptotic limit of $\theta$ as in Equation~\eqref{krechmanneq}, we obtain:
\bea
\tilde{c}L_p^{\frac{3}{2}}=\frac{3}{\sqrt{\pi}}\sqrt{2GM}.
\eea
Thus, $F(r)$ in Ansatz~\eqref{Schilsmetric} takes the following form:
\bea
F(r)=\frac{18GM}{r^4}(\int_0^rdy\sqrt{y}\int_0^{\infty}\frac{\tilde\mu (dw)}{w^{\frac{3}{2}}}\int_0^{\frac{yw}{L_p}}e^{-z}z^{\frac{1}{2}}dz)^2.
\eea
Furthermore, the area $A$ in Equation~\eqref{qmSurface} simplifies to:
\bea
A=\frac{8\pi^2}{9}\int_0^{r_h}\tilde{B}(x)dx,
\eea 
where $\tilde B(x)$ is defined by:
\bea
\tilde{B}(x)=x^{4}(\int_0^xdy\sqrt{y}\int_0^{\infty}\frac{\tilde\mu (dw)}{w^{\frac{3}{2}}}\int_0^{\frac{yw}{L_p}}e^{-z}z^{\frac{1}{2}}dz)^{-2}.
\eea
Thus, the entropy defined by $S=\frac{A}{4G}$ is again greater than $\frac{\pi{r_h}^2}{G}$. Therefore, we propose the following conjecture:
\begin{itemize}
	\label{boundconjecture}
	\item The Entropy Bound Conjecture: The true quantum Schwarzschild black hole always have an entropy satisfying $S>\frac{A_h}{4G}$,~where $A_h$ is the area of the horizon.
\end{itemize}

\subsection{Temperature-Mass Diagram}
Regular black holes exhibit thermodynamic properties that differ significantly from those of the classical Schwarzschild black hole. In the classical framework, the temperature of a black hole is given by:
\bea
T=\frac{1}{8\pi G M}.
\eea 
This demonstrates that the temperature of a black hole increases unboundedly during Hawking radiation, rendering the black hole unstable under this process. However, for quantum regular black holes, this is not the case, as their temperature is bounded by the Planck temperature. This behavior is evident in the function $F(r)$ in Figure~\ref{diffrentPlancklength}, which shows a critical point at $r = r_c$, where $F(r_c) = 1$ and $|F'(r)|_{r = r_c} = 0$. This indicates that the temperature of the black hole becomes zero at this critical point. We expect that for large $M$, the classical results remain a good approximation. Therefore, for regular black holes, there exists a maximum temperature that roughly scales as $T \sim \frac{1}{L_p}$. For black holes described by the ansatz with $F(r)$ given in \eqref{testfunction}, the temperature-mass diagram in the Planck units, with $L_p = 1$, is shown in Figure~\ref{mass-tem}:
\begin{figure}[H]
	\centering
	\includegraphics[width=0.57\linewidth]{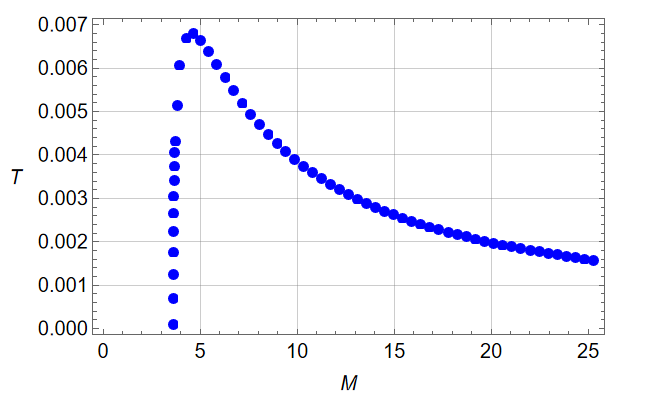}
	\caption{Temperature-Mass diagram for $L_p=1$. }
	\label{mass-tem}
\end{figure}
As shown in this figure, the temperature of the black hole is zero at $M=3.633$, and the maximum value of the temperature occurs just above this mass scale. From the temperature-mass diagram, it is also evident that the classical limit $T=\frac{1}{8\pi L_p^2M}$ cannot be achieved by a continuous or perturbative deformation of the corresponding regular black holes. Suppose that $M^{*}$ is the critical mass for the maximum temperature, then the specific heat $\frac{1}{C_V}\sim\frac{dT}{dM}$ is positive for $M<M^{*}$ and negative for $M>M^{*}$. In the classical limit $L_p\to0$, $\frac{dT}{dM}\to\infty$ for $M\to M^{*-}$ and $\frac{dT}{dM}\to-\infty$ for $M\to M^{*+}$\footnote{We denote $M^{*\pm}$ as left-hand and right-hand limits at $M=M^{*}$.}. The process indicates that if the classical limit $L_P\to0$ exists, then it is not a smooth process.
 
\subsection{Some Comments on More Generic Spherically Symmetric Black Holes}
The preceding results are based on Ansatz~\eqref{schwartcoordinate}. However, there is no a priori reason for spherically symmetric ansätze to satisfy $g_{tt}g_{rr}=-1$ in quantum gravity. Furthermore, quantum corrections to the Schwarzschild black hole~\eqref{quantumSch} suggest that this is not the case. Therefore, we focus on a more general spherically symmetric ansatz as follows: 
\bea
\label{genericspherically}
ds^2=-(1-\bar{F}(r))dt^2+\frac{1}{H(r)(1-\bar{F}(r))}dr^2+r^2d\Omega^2,
\eea
where we assume that $H(r)$ is a positive bounded function over $[0,\infty)$, and also $g_{tt}$ and $g^{rr}$ share a common zero.

The expansion scalar for marginally bound timelike geodesics is given by: 
\bea
\label{generictheta}
\theta=\frac{\sqrt{H}}{r^2}(r^2\sqrt{\bar{F}})'.
\eea
Since we have assumed that $H(r)$ is a non-zero bounded function, the regularity of $\theta$ implies that the near-center behavior of $\bar{F}(r)$ is the same as before:
 \bea
 \label{ansatz2}
 \bar{F}(r)\sim r^2 +O(r^3).
 \eea
Additionally, the asymptotic behaviors of the functions $H(r)$ and $\bar{F}(r)$ as $r\to\infty$ must be consistent with the Schwarzschild spacetime; thus, we have:
\bea
\lim_{r\to\infty}\bar{F}(r)=\frac{2M}{r},~~~~\lim_{r\to\infty}H(r)=1.
\eea
The Ricci scalar corresponding to this ansatz is:
\bea
R=\frac{4+r(-4+4\bar{F}+r\bar{F}')H'+2H(-2+2\bar{F}+4r\bar{F}'+r^2\bar{F}^{''})}{2r^2}.
\eea
Using the near-center approximation in Equation~\eqref{ansatz2},~$\bar{F}(r)=a_1r^2+a_2r^3+...$, we obtain:
\bea
R=2\frac{1-rH'-H}{r^2}+3a_1rH'+12a_1H+... .
\eea
The regularity of the first term implies that $H(r)=1+b_1r+b_2r^2+...$ as $r\to 0$. Therefore, $H(r)$ approaches unity in both asymptotic limits~$r\to 0$ and $r\to\infty$. Using these asymptotic limits and similar reasoning, we conclude that singularities can be avoided only through non-perturbative corrections in~$\hbar$ within quantum gravity. 

To determine $\bar{F}(r)$ and $H(r)$, we need two completely monotonic functions. The Kretschmann scalar for Ansatz~\eqref{genericspherically} is:
\bea
\label{kretchmanneqs}
K(r)&=&\frac{1}{4r^4}\Big(16+r^2(8(1-\bar{F})^2+r^2\bar{F}^{'2})H^{'2}+4H(r^4\bar{F}'\bar{F}^{''}H'-4(1-\bar{F})(2+r^2\bar{F}'H'))\nn\\
&+&4H^2(4+4(-2+\bar{F})\bar{F}+4r^2\bar{F}^{'2}+r^4\bar{F}^{''2})\Big).
\eea
The asymptotic limit of the Kretschmann scalar is $lim_{r\to\infty}K(r)=\frac{48M^2}{r^6}$. The regularity of the Kretschmann scalar at $r=0$ implies that $b_1=0$ in the expansion of $H(r)$. Furthermore, the complete monotonicity of $K(r)$ implies that $\Big(5a_2a_3-2(a_3b_2+a_2b_3+b_2b_3)\Big) <0$. Evidently, obtaining an analytical solution for the coupled differential equations~\eqref{kretchmanneqs}~and~\eqref{generictheta} is infeasible. Since $H(r)$ is assumed to be positive over $[0,\infty)$, the temperature of the black hole described by Equation~\eqref{genericspherically} is:
\bea
T=\frac{1}{4\pi}\sqrt{H(r_h)}|\frac{d\bar{F}}{dr}|_{r=r_h},
\eea
where the horizon is again the solution to $\bar{F}(r_h)=1$. By choosing $\theta(r)$ as in Equation~\eqref{qtheta}, we obtain:
\bea
T=\frac{1}{4\pi}\frac{dlog(\bar{B}(r_h))}{dr_h}\sqrt{H(r_h)},
\eea
where $\bar{B}(r_h)$ is given by:
\bea
\label{integral}
\bar{B}(r_h)=r_h^4\Big(\int_0^{r_h}dy\frac{\sqrt{y}}{\sqrt{H(y,M(y))}}\int_0^{\frac{y}{L_p}}e^{-z}z^{\frac{1}{2}}dz\Big)^{-2}.
\eea
The function $H(r)$ may also depend on the mass; thus, this dependence must be carefully considered. The solution to $\bar{F}(r_h)=1$ can be expressed as $M=M(r_h)$, where the function $M(r_h)$ represents the mass which depends of the horizon length. This dependence must be considered in $H(x)$ in the integral in Equation~\eqref{integral}; thus, we express $H(x)$ as $H(x,M(x))$. By straightforward calculations, the entropy of the black hole is $S=\frac{A}{4G}$, where $A$ is given by:
\bea
A&=&\frac{8\pi^2}{9}\int_0^{r_h}\frac{\bar{B}(x)}{\sqrt{H(x,M(x))}}dx\nn\\
&=&\frac{8\pi^2}{9}\int_0^{r_h}\frac{x^4}{\Big(\int_0^xdy\frac{\sqrt{y}}{\sqrt{H(y,M(y))}}\int_0^{\frac{y}{L_p}}e^{-z}z^{\frac{1}{2}}dz\Big)^{2}\sqrt{H(x,M(x))}}dx.
\eea
This area satisfies the following bound:
\bea
A\geq\frac{32\pi}{9}\int_0^{r_h}\frac{x^4}{\sqrt{H(x,M(x))}(\int_0^x\sqrt{\frac{y}{H(y,M(y))}}dy)^2}dx.
\eea

Conjecture~\ref{boundconjecture} depends significantly on the function $H(r)$, which satisfies Equation~\eqref{generictheta} and Equation~\eqref{kretchmanneqs} for completely monotonic functions $K(r)$ and $\theta(r)$. Notably, the entropy bound is satisfied for certain regular black holes within this broader class of black holes, as explored through the loop quantum gravity approach~\cite{belfaqih2024blackholeseffectiveloop}.

It would be intriguing to compare this conjecture with the entropy bound for a broader class of black holes. We leave this question as a promising avenue for future research.

\section{Conclusion}
In this paper, we have explored a class of regular black holes for static, spherically symmetric spacetimes that enjoy complete monotonicity. We indicated that complete monotonicity may represent a fundamental characteristic of quantum black holes, at least for certain curvature invariants such as the Kretschmann scalar and the expansion scalar for marginally bound timelike geodesics. This has been done by introducing suitable completely monotonic functions with a bounded measure in Bernstein's theorem. It is shown that the boundedness of measures is equivalent to the regularity of the ansatz.  Also, we have found a bijection map between this class of functions and the all completely monotonic functions.

The same completely monotonic behavior appears in some models of electrodynamics such as the corrected Coloumb potential at one loop in QED, the regularized Maxwell electrodynamics, and ModMax theory. However, some models superficially show some non-monotonicity, but as we argued, this violation of complete monotonicity is not reliable, since perturbation theory breaks down in regimes where non-monotonicity occurs. This leads to the non-perturbativity conjecture, which states that complete monotonicity cannot be falsified by perturbative corrections.

As argued, for static, spherically symmetric spacetimes satisfying $g_{tt}g_{rr}=-1$, the singularity of black holes cannot be resolved by perturbative corrections to the Einstein-Hilbert action, and some non-perturbative corrections should be considered. We also show that it is true for a more general black hole ansatz as given in Equation~\eqref{genericspherically}.

We have also studied the thermodynamics of such black holes and provided some evidence suggesting that the entropy of black holes described by Ansatz~\eqref{schwartcoordinate} satisfies $S>\frac{A_h}{4G}$, where $A_h$ is the area of the horizon. It leads to the entropy bound conjecture, stating that the entropy of quantum regular black holes is always greater than $\frac{A_h}{4G}$. We also demonstrated that these black holes exhibit a bounded temperature, in contrast to the unbounded temperature of classical Schwarzschild black holes, with a maximum temperature scaling as $T\sim\frac{1}{L_p}$ and a critical mass scale where the temperature vanishes, indicating a stable endpoint for Hawking radiation.

Finally, we showed that non-perturbative corrections to the Einstein-Hilbert action are necessary to avoid singularities in a more general class of static, spherically symmetric spacetimes, as in Ansatz~\eqref{genericspherically}. However, for this class of backgrounds, the conjectured bound on the entropy of black holes cannot be tested by analytic approaches and remains a topic for future research.
\section*{Acknowledgments}
The authors thank Hessamaddin Arfaei and Martin Bojowald for their valuable comments. A. Moradpouri also extends his gratitude to Mohammad Javad Tabatabaei and Mohammad Ebrahimi for their assistance with numerical computations.
 
\bibliographystyle{JHEP}
\bibliography{mybib}  
\end{document}